# A Novel Approach for Establishing Connectivity in Partitioned Mobile Sensor Networks Using Beamforming Techniques

**Abbas Mirzaei***
Department of Computer Engineering, Ardabil Branch, Islamic Azad University, Ardabil, Iran
a.mirzaei@iauardabil.ac.ir
**Shahram Zandiyan**
Department of Computer Engineering, Ardabil Branch, Islamic Azad University, Ardabil, Iran
Shahram.zandian.8872@gmail.com

Abstract
Network connectivity is one of the major design issues in the context of mobile sensor networks. Due to diverse communication patterns, some nodes lying in high-traffic zones may consume more energy and eventually die out resulting in network partitioning. This phenomenon may deprive a large number of alive nodes of sending their important time critical data to the sink. The application of data caching in mobile sensor networks is exponentially increasing as a high-speed data storage layer. This paper presents a deep learning-based beamforming approach to find the optimal transmission strategies for cache-enabled backhaul networks. In the proposed scheme, the sensor nodes in isolated partitions work together to form a directional beam which significantly increases their overall communication range to reach out a distant relay node connected to the main part of the network. The proposed methodology of cooperative beamforming-based partition connectivity works efficiently if an isolated cluster gets partitioned with a favorably large number of nodes. We also present a new cross-layer method for link cost that makes a balance between the energy used by the relay. By directly adding the accessible auxiliary nodes to the set of routing links, the algorithm chooses paths which provide maximum dynamic beamforming usage for the intermediate nodes. The proposed approach is then evaluated through simulation results. The simulation results show that the proposed mechanism achieves up to 30% energy consumption reduction through beamforming as partition healing in addition to guarantee user throughput.

## 1- Introduction

Mobile sensor networks (MSN) are effective platforms for the industrial and military communications which are applied for commercial affairs of the manufacturing sector and the identification of enemy frontiers in the military. In most applications, sensors have the role of a data source and send information from event triggers to each eNodeB or central receiver. The unique role of the base station makes it a natural target for the enemies who intend to carry out the deadliest attack with the least possible effort against the mobile sensor network. Even if the mobile sensor network uses common security mechanisms such as encryption and authentication, the enemy may use traffic analysis techniques to identify the base station. However, the attractiveness of mobile sensor networks and their advantages make them vulnerable to potential attack by the evil enemy. A typical mobile sensor network consists of several relays that iteratively transfer new information to the existing BS. In this model, because the unique role of the base station makes it possible to carry out the most effective attack against the target mobile sensor network with the least possible effort, this station becomes the center of enemy attacks. That is, the enemy assumes that a Denial of Service (DoS) attack against the base station will actually cripple the larger mobile sensor network, because the base station not only acts as a data well.

One of the main effective ways to protect a base station from a vicious enemy attack is to keep its role, identity, and location unknown. However, conventional security mechanisms that provide confidentiality, integrity, and authentication are not capable of this type of protection [1] [2]. One of the major portions of the studies relevant to anonymous communications were so far related to analyzing routing algorithms with the aim of concealing actual paths from the transmitter well [3], [4]. It should be noted that, in spite of the fact that secure routing algorithms can greatly reduce path discovery attack, the enemy can gain important data via monitoring the link layer and the relevance between pairs of nodes, based on which it can identify the location and role of the base station [5] [6].

According to [7] [8], the authors suggested an approach in the lower layer which uses dynamic beamforming to further identify the base station. Nowadays, distributed beamforming seems a very attractive way to improve the network performance, throughput and power utility, provide data link safety, in addition to increasing SINR in the multi-layer cooperative systems [9] [10]. Based on dynamic beam modulation, several mobile sensor network nodes work together to share the existing propagation capabilities in order to create a dynamic multiple transmission network. Various relays are able to concurrently transmit information, taking into account the conditions of the wireless channel and the precise control of the signal phase, in such a way that all the signals are combined at the destination. For example, ideally, N transmitters send the same messages with the same power, while tolerating a path loss during the transmission of the signal to a normal destination increases the power at the destination by N times. This feature has been shown to increase base station anonymity in mobile sensor networks. This protocol appropriately disrupts the evidence hypothesis (EH) and it doesn't consider the real base station of the mobile sensor network well.

This protocol is an effective technique for enhancing the probability of low-cost, multi-hop paths usage, and the power needed to transmit the signal to the destination is used as the cost of the L-link. Because, the average energy consumption cost of the protocol increases with increasing the number of auxiliary relays |L|, the use of L link selecting the paths to maximize |L| can increase the base station anonymity with energy costs equal to the mobile sensor networks with anonymous protection [11].

As far as we know, participatory communication was first used to reinforce the base station anonymity in Ref. [12]. As a result, the former studies of distributed beam formation and base station anonymity will be discussed separately. Researchers on the subject of base station anonymity initially defined a quantitative way of measuring anonymity. Some researchers developed sub-optimal effective approaches for measuring anonymity in the connection entropy [13], GSAT test [14], and belief [15] [16]. Entropy and GSAT methods impose certain limitations on the enemy. They give the a priori possibility that the location of the base station is known to the enemy or that the enemy can estimate the location of the base station. The functionality of the belief index, according to the evidence theory, does not have any of these hypotheses and so it has attracted a lot of attention as a metric for recognizing anonymity. In Section 4, we discuss the evidence theory and the metric of belief to evaluate base station security. Many published techniques for dealing with the traffic insecurity in mobile sensor networks applied various approaches to make the location of data sources hidden [17] [18]. Such as [19] in which the authors propose various approaches such as uniform packet speed and false paths to confuse the enemy. Similarly, the authors in [20] suggest that network paths be modified by considering virtual sinks. Two techniques have been proposed in [21]. In the first technique, the base station re-transmits a package of received packets at various degrees and the base station looks like an ordinary node for the enemy. The base station can also be considered and can move to a safer location.

The above techniques are used in the network layer of communication protocols. The protocol uses distributed beamforming in the physical layer to improve the base station anonymity [22]. This paper compensates for this shortcoming of the previous protocol by providing a multi-layered routing algorithm considering data link constraints and auxiliary intermediate nodes in order to decrease the total power utilization.

The authors in [23] proposed an efficient smart control plan for the dynamic transmission in the wireless sensor networks using cooperative protocols. This algorithm supports the dynamic operations of block data and third-party public validation to provide high security against data forgery and replacement. In [24], a QoS model for resource allocation algorithm was proposed for data replicas based on the servers existing in a network in order to improve the connectivity approach service and decrease total cost. In [25], a distributed algorithm was proposed to reduce the access delay and expand the network bandwidth. In this scheme, a new data analysis strategy was proposed to mitigate the costs of data storage and information transfer for applications. In [26], a close-loop content-oriented scheme was reviewed achieving higher performance for data-intensive applications. Also, some researchers addressed main critical challenges of this criterion, such as energy efficiency [27] availability [28], and security [29] of data access. However, heterogeneous MSN has security challenges, including vulnerability for sensors and association acknowledgment, that delay the rapid adoption of computing models.

Unfortunately, the abovementioned works cannot be considered as a proper approach for large-sized networks due to reliability conditions and high computational complexity at the central unit that significantly increases to the number of sensors in the network.

In this paper, we present the cost of the L link, which has been optimized for multi-hop paths that minimize the average power consumption of the mobile sensor network. Using simulations, we show that the cost of our link is such that it maintains the anonymity of the base station while reducing the communication energy consumption. This article continues as follows: Section 2 examines the system model and the problem formulation. Section 3 provides a framework for Distributed beamforming and the energy efficiency of the protocol. Section 4 describes the proposed approach in Power Optimization in mobile sensor networks. The numerical results and discussion of the proposed approach are presented in Section 5. Finally, Section 6 draws the conclusion and highlights future challenges to motivate the effective integration of beamforming-based mobile sensor networks with the diagnosis.

## 2- System Model and Problem Formulation

### A- Network model

In this paper, we consider a homogeneous model for a mobile sensor network in which all sensor nodes have the same capabilities in terms of battery life, type of radio communication, and network protocols. In this paper the sensors were considered as mobile nodes. The base station acts as a well for all data traffic generated by the sensor nodes. There is only one base station on the network. Our hypothesis is that the sensors can be aware about the locations of the base station and the neighbor sensors as well [30]. In addition, the cells are well-informed about the level of transmission energy needed to get all subsequent hops. Multi hop paths are followed to deliver data frames to the base station. Also, the cross-wave propagation model has been considered in this paper.

We assume that precautions are used in the design and operation of the base station to prevent enemy infiltration. For example, the base station maintains the transmission power level equal to the other cells (for example, updated path exploration and authentication messages) so that it cannot be detected from other sensor nodes by radio frequency analysis. Messages are transmitted with the header and encrypted message body. We assume that the TDMA Media Access Control Protocol (MAC) operates by synchronizing sufficient time on all wireless network sensors at tolerable shielding intervals [31]. All nodes in mobile sensor networks are considered as auxiliary relay options.

**B-Problem Formulation**
Assume that the mobile sensor network transmits the target sensitive data, which is a desirable target for the enemy. After identifying and abusing the base station, the enemy aims to carry out a DoS attack against the base station at any cost, such as physically destroying the base station. Also, the enemy is actively engaged in eavesdropping by being present in all parts of the mobile sensor network [32] [33]. The enemy is able to identify the location of all radio communications at the location of the network [34]. While the enemy monitors the traffic, we assume that the cryptographic system is robust enough so that the enemy cannot use the cryptographic system analysis to retrieve the contents of the body or header. The enemy uses the evidence theory traffic analysis to localize the base station, unaware that the mobile sensor network is using the distributed beamforming.

The enemy starts via monitoring the transmit links demonstrated by $E(U)$ in which U is a direct connection among the nodes ($S_i$ & $D_i$). It also obtains the paths by correlating all the evidence for the node pair. The overall path containing two or more nodes is denoted by V, and the associated evidence $E(V)$ is calculated as follows:

$$E(V) = \min_{U \subseteq V}\{E(U)\}, \qquad |V| \geq 2, \qquad (1)$$

Normalized evidence $m(V) = E(V)/\Sigma E(V)$ shows a proportion of all the evidence gathered by the enemy that supports the $B(u)$ which indicates the enemy's certainty that there is a path of length n in any given node and is expressed as follows:

$$B(u) = \sum_{U|u \subseteq V} n\, m\,(U). \qquad (2)$$

In this paper, the Belief index has been applied in order to evaluate base station anonymity. The small belief metric means less confidence of the enemy or more anonymity of the base station coordination. To reduce the computational complexity of the calculations needed, it's supposed that the enemy splits the mobile sensor network into an $M \times M$ network consisting of $N_C$ square cells. This means that the enemy only needs to identify the target location within the cell. As a result, the belief metrics $B(u)$ generated by cell analysis cause u to indicate that the sensor is not a specific sensor, but one of the $N_C$ cells in the enemy's target network. Section 5.b has presented an example of evidence theory analysis. Figure 1 illustrates the network configuration and the communication beamforming links between the sensors and the base station.

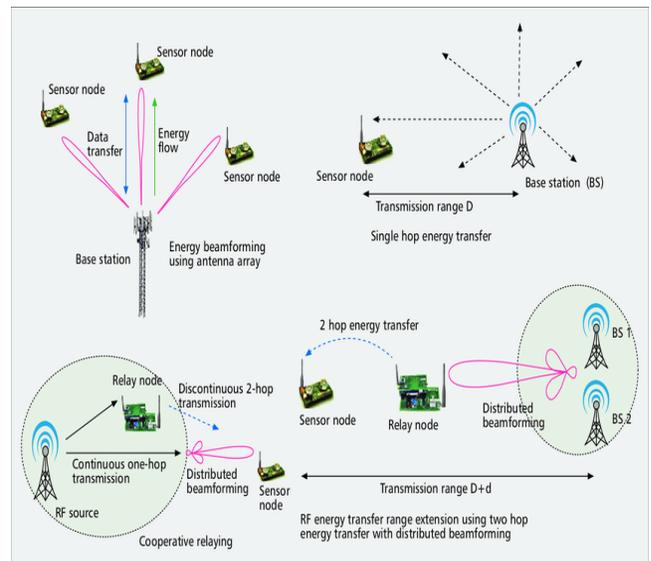

Fig. 1. The network configuration and the beamforming between sensors and the base station.

## 3- Distributed Beamforming Model

### A. Distributed Beamforming
Based on the proposed methodology, a distributed beamforming approach has been applied to further identify the base station. Distributed beamforming uses the broadcast nature of wireless transmission. Adjacent nodes may also hear all the frames sent to a particular receiver. According to Figure 2, these adjacent nodes may act as auxiliary relays in cooperation with the transmission source $S_i$ so that the transmitted signal travels to $D_i$ through a diverse set of transmitters. Because each $R_j$ relay sends the same message $S_i$ with the exact time and synchronization of the carrier, the signals at the destination $D_i$ are combined under the conditions of ideal scheduling and carrier synchronization [35].

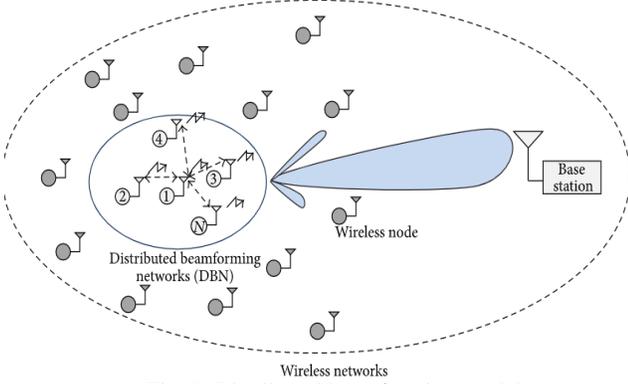

Fig. 2. Distributed beamforming model

Ideal received signal in $D_i$ by source $S_i$ and $|L|$ auxiliary relay $R_j$ is sent as shown below:

$$r_{D_i}(t) \triangleq r_{S_iD_i}(t) + \sum_{j=0}^{|L|-1} r_{R_jD_i}(t)$$

$$= \Re\left(A_{S_i}(t)w_{S_iD_i}\beta h_{S_iD_i}e^{j(2\pi f_c t+\theta(t)+\varphi(t))}\right) +$$

$$\Re\left(\sum_{j=0}^{|L|-1} A_{R_j}(t)w_{S_iD_i}\beta h_{S_iD_i}e^{j(2\pi f_c t+\theta(t)+\varphi(t))}\right) + n(t) \quad (3)$$

In this equation, $h$ indicates the channel shock response, $\beta$ indicates the sharing efficiency $f_c$ is the carrier frequency, $\theta(t)$ illustrates the phase modulation expression, $\varphi(t)$ is total phase variation term and $n(t)$ represents the thermal noise contained in the $D_i$ receiver. $r_{D_i}(t)$ consists of two expressions: Information received from source $S_i$ (i.e., $r_{S_iD_i}(t)$) and total information received from the relay $|L|$ used $R_j \in L$ is equal to $\sum_{j=0}^{|L|-1} r_{R_jD_i}(t)$.

The meaning of Equation 3 in terms of achieving base station anonymity in the physical layer is that the distribution of the distributed beamforming makes it possible for the component of the signal received $r_{S_iD_i}(t)$ from the transmitted information $S_i$ decreases by $\sum_{j=0}^{|L|-1} r_{R_jD_i}(t)$ at the same time, the power level of the signal received by $D_i$ remains constant during the phase offset $\varphi(t)$. Therefore, if $S_i$ and $R_j\epsilon L$ transmit only data at a specified SINR with the power required to reach $D_i$, each transmitter can reduce the resource via $10log(|L|+1)$ dB and properly prevents the enemy from distinguishing $E(S_i, D_i)$. Accordingly, the elimination of $E(S_i, D_i)$ from the enemy evidence set increases the confidentiality of the base station, as it reduces its role in detecting $B(u = D_i)$.

Figure 2 shows the functionality of the distributed beamforming used in any relay via the node $S_i$, which intends to apply dynamic beamforming to increase the base station anonymity while sending a packet to the next relay node $D_i$. $S_i$ should first choose the appropriate subset of auxiliary nodes by handshaking (according to steps (a) & (b)). When a vector of auxiliary relays is used, $S_i$ transfers the packet body in step $d$ to $R_j \in L$. In step $e$, the distributed beamforming is submitted and then the authentication message is transmitted from node $D_i$ to $S_i$ in stage (f) to confirm the correct reception of the cooperative message.

## B. Distributed beamforming protocol

Figure 3 is an example of an enemy theory (ET) analysis with and without using distributed beamforming for seven network sensors. In this example, the enemy divides the target area into $N_c = 9$ cells. In Figure 3, the transmission of information in the main method along with the cooperative transmission by the distributed beamforming. In continuation, the paper shows the evidence collected and the belief calculation in unit hops in the main method and the distributed beamforming protocol, in which the relay in cluster 28 sends the packet to the relay in cluster 35.

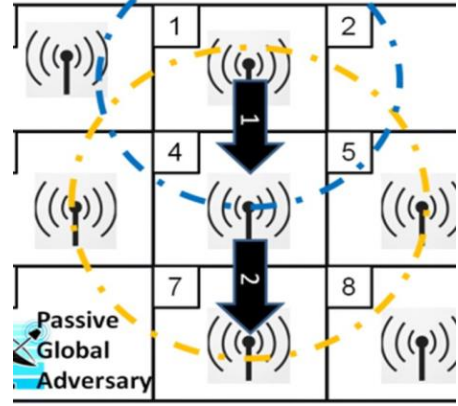

Fig. 3. Applying enemy-theory for a mobile sensor network

## C- Distributed beamforming energy analysis

In addition to the higher anonymity of the base station, the energy cost of communications resulting from the use of the distributed beamforming protocol should also be considered. While it may save up to $10Log(|L+1|)$ on the transmission power, additional signaling in the distributed beamforming protocol causes increase of the data overhead. Dedicated power for each sent bit $\varepsilon_b$ can be computed by ratio of the mean transmit energy $P_T^{\overline{S_iD_i}}$ needed to reach the $S_i$ to $D_i$ information by the signal to noise ratio (SNR) and the set speed $r$, where $\overline{\varepsilon_b^{S_iD_i}} \triangleq \frac{P_T^{\overline{S_iD_i}}}{r}$. Therefore, the mean communication energy consumed by the non-cooperative system is equal to $\overline{\varepsilon_{base}} \triangleq \overline{\varepsilon_b^{S_iD_i}}\overline{\beta}$, where $\overline{\beta}$ is equal to the average body size. Accordingly, the mean distributed beamforming power is calculated based on a cooperative hop and is obtained based on the following formula:

$$\overline{\varepsilon_{DIBAN}} \triangleq \left( \varepsilon_b^{\overline{S_iR_j}}(\gamma_{RR} + \gamma_{data} + \bar{\beta} + K) + \frac{\varepsilon_b^{\overline{S_iD_i}}}{(|L|+1)}\bar{\beta} + \right.$$
$$\left. |L|\varepsilon_b^{\overline{R_jS_i}}\gamma_{Ack} + |L|\frac{\varepsilon_b^{\overline{R_jD_i}}}{(|L|+1)}\bar{\beta} + \varepsilon_b^{\overline{D_iS_i}}\gamma_{Ack} \right), \qquad (4)$$

Where $Y_{RR}$ represents the Relay Request, $Y_{Data}$ denotes the Data Multihop and $Y_{Ack}$ is equal to the header size of acknowledgement/negative acknowledgement.

In this scenario, the channel was considered to be equal to $K$ bits at time $t$ microseconds (i.e., $K = t \times r$) [7].

In equation (4), we have two key observations of the average the distributed beamforming energy consumption per hop ($\overline{\varepsilon_{DiBAN}}$). First, $S_i$ does not make saving on the distributed beamforming for power transmission ($\overline{P_T^{S_iR_j}}$) which involves the use of relays. As a result, $S_i$ uses relays that are as close as possible to $S_i$, thus minimizing ($\varepsilon_b^{\overline{S_iR_j}}$). Second, $S_i$ seeks to maximize $|L|$ by applying the maximum number of auxiliary relays possible and at the same time minimizing the use of transmission power ($\overline{P_T^{S_iR_j}}$).

### D- Selection of relay in the distributed beamforming protocol

The distributed beamforming protocol requires an approach to select a relay in order to use a set of $R_j \in L$ in each hop and has three objectives: higher anonymity of the base station, conservation or reduction of communication energy $\overline{\varepsilon_{DiBAN}}$ compared to the main system $\overline{\varepsilon_{base}}$ without distributed beamforming and achieving the best CSI measurement. Increased | L | causes the reduction of the ability to communicate directly with $S_i$ and $D_i$ and thus improves the base station anonymity. But as we said before in the previous section, the increase of $|L|$ requires higher transmission power ($\overline{P_T^{S_iR_j}}$) to be used during relay operation. By iteration, we obtain the expected number of potential relays that $S_i$ made available by increasing the power level ($\overline{P_T^{S_iR_j}}$). Given the constraint $P_T^{S_iR_j} < \overline{P_T^{S_iR_j}}$, we maintain the anonymity of the base station to prevent the enemy from collecting the evidence of $E(S_i, D_i)$ linking the transmitter and receiver.

First, the quantity of the potential intermediate nodes was considered equal to $|L_D| = \lambda(\frac{\pi d^2 S_iR_j}{8})$ where $\lambda$ is the density of the node in the region, which is $\lambda = \frac{S_U}{M \times M}$ in a semicircle with radius $d_{S_iR_j} = \frac{d_{S_iD_i}}{\delta}$ is calculated based on the number of expected nodes in the receiving interval when $S_i$ is transmitted with power $P_T^{S_iD_i}$. This algorithm adjusts $P_T^{S_iR_j}$ using $\delta$ through iteration where $\delta > 1$, because $\delta = 1$

means that $P_T^{S_iR_j} = P_T^{S_iD_i}$ and there is an undesirable link between $S_i$ and $D_i$.

Our relay selection algorithm is briefly performed based on the below stages:

1- $S_i$ chooses the primary amount of $\delta$ and mathematically calculate $|L_D|$. In the beginning, the set of the distributed beamforming relays is $L = \emptyset$.

2- $S_i$ sends the relay request message and $P_T^{S_iR_j}$ is calculated based on the $\delta$ to reach $|L_D|$.

3- Nodes that respond to $S_i$ with a confirmation message involves the number of available relays $|L_A|$ in a way that $R_j \varepsilon L_A$.

4. One of the following three results occurs for $R_j \in L_A$:

A- If $|L_A| < |L_D|$, then $\delta = \delta - \delta_{STEP}$ is used to increase $P_T^{S_iR_j}$ and reach more candidate relays in each iteration. Return to step 2. If $\delta = min(\delta)$, the algorithm ends with $L = \emptyset$ and the distributed beamforming is not used in this hop.

B- If $|L_A| = |L_D|$ , then $L = L_A$ and the algorithm terminates.

C- If $|L_A| \geq |L_D|$, L becomes $|L_D|$ of the relay of the highest quality $R_j \in L_A$ which are prioritized based on the best condition of channel state information.

The proposed node selection approach is used at all relays next to the transmitter path to the base station. Decreasing $\delta$ is the only node solution to use more intermediate nodes to enhance $\overline{P_T^{S_iR_j}}$ . Nevertheless, the multi-layer routing approach offers another option to increase | L | for the mobile sensor network: In this case, the paths are selected as a functionality of the accessible auxiliary intermediate nodes $|L_A|$ in all hops.

The cost of a $\mathcal{L}_i$ link consists of two parameters, each with a unique objective for the distributed beamforming. First, energy saving is the most important factor in the initial design of mobile sensor networks and is one of the fundamental technical design constraints for further anonymity of the base station. Instantaneous energy required to use a set of the auxiliary relay $|L_A|$ is equal to:

$$\varepsilon_{RR} \triangleq \sum_n \left( \left( \varepsilon_b^{S_iR_j} \times \gamma_{RR} \right) + \gamma_{Ack} \sum_{j=0}^{|L_A|-1} \varepsilon_b^{R_jS_i} \right), \qquad (5)$$

The reader is reminded that while selecting a relay, the energy to bit $\varepsilon_b^{R_jR_j}$ required to use $|L_A|$ the relay is dependent on $\delta$. quantity of re-transfers needed for successful use of a suitable set of $|L_A|$ of the relay, a relay in which the relay selection condition is met is represented by $n$. Secondly, the average energy consumption of the distributed beamforming decreases with increasing the number of $|L_A|$ of the auxiliary relays available in each hop. As a result, the cost of the beam links is calculated based on the following formula:

$$\mathcal{L}_i = \begin{cases} \left(\dfrac{\varepsilon_{RR}}{|L_A|}\right) for \ |L_A| > 0 \\ \infty \qquad for \ |L_A| = 0 \end{cases} \qquad (6)$$

When the auxiliary relays are available, the cost of link $\mathcal{L}_i$ is decimal and when the relays are not available it becomes $\infty$. The cross-layer scheme integrates with the distributed beamforming relay selection algorithm so that each node can calculate $|L_A|$ and make better use of the distributed beamforming by selecting routes with higher relay densities.

# 4. Power Optimization in Mobile Sensor Network

According to the main sources in this field, there is no standard model for energy consumption in beamforming-based mobile sensor backhaul networks. However, the application of nonlinear prediction energy consumption in such systems has attracted more satisfaction. Here, this paper uses adaptive resource allocation in which the backhaul connection has been modeled, in which C5 and C6 are the maximum transmission power constraints for sensors and macro base stations, respectively.

## A. Content-Caching model

In this network, we suppose that content can be modelled as a distinct set of packet data as $\mathcal{F} = \{\mathcal{F}_1, \mathcal{F}_2, ..., \mathcal{F}_f, ..., \mathcal{F}_F\}$ which $\mathcal{F}_f$ represents the $f$-th data frame. The request probability for data frame $f$ is expressed as

$$p_f(0 \leq p_f \leq 1), \qquad which, \sum_{f=1}^{F} p_f \leq L_i, \ \forall f \in \mathcal{F}, \quad (7)$$

It should be noted that the caching model presented for this paper is stochastic caching so that we can calculate the probability of caching data packet $f$ via base station $i$ $0 \leq q_{f_i} \leq 1$ where $L_i$ illustrates the cache size. In addition, $\{q_{f_i}\}$ of base station $i$ should satisfy below condition:

$$\sum_{f=1}^{F} q_{f_i} \leq L_i, \ \forall i \in \mathcal{B}, f \in \mathcal{F}, \qquad (8)$$

$L_i = L_M$ means the cell is macro cell, otherwise, $L_i = L_S$.

## B. Resource Control Model

Based on the approach's principles, the resources of the base stations can be supplied by conventional smart grid and renewable energy harvesting. During this scenario, the transmission power relevant to base station $i$ can be demonstrated by $P_i(i \in B)$, and the applied energy from the grid network is illustrated by $G_i$. The harvested renewable resource is shown as $E_i$. Based on the enabled power sharing capability, the shared power among cell $i$ and cell $i'$ is equal to $\varepsilon_{ii'}$, where $\beta \in [0,1]$ denotes the power-sharing index among base stations. So, we can conclude that $(1 - \bar{\beta})$ is equal to the loss percentage in the power sharing stage. The following condition should be satisfied during the power sharing process.

$$P_i < G_i + E_i + \beta \sum_{i' \in \mathcal{B}, i' \neq i} \varepsilon_{i'i} - \sum_{i' \in \mathcal{B}, i' \neq i} \varepsilon_{ii'}. \qquad (9)$$

According to the defined conditions, the overall power efficiency can be affected by the transmission strategies, power sharing and the level of harvested energy from renewable resources.

## C. Transmission Model

Taking fairness into account, we tried to provide data rate balancing throughout the network. In which, $x_i(i \in \mathcal{B}, j \in \mathcal{U})$ denotes the association indicator, for example, $x_{ij} = 1$ represents that node $j$ is associated with BS $i$ and otherwise the node has not been associated with the base station. Subsequently, $k_i = \sum_{j \in \mathcal{U}} x_{ij}$ represents the number of sensors associated to cell $i$. $\left(\sum_{f=1}^{F} p_f q_{f_i}\right)^{k_i}$ express the probability of serving $k_i$ associated nodes by base station $i$. if $x_{ij} = 1$, we can calculate the efficiency of the $j$-th sensors as $\mu_{ij} = \log(R_{ij})$ which $R_{ij}$ is the throughput so that the $R_{ij}$ is obtainable as.

$$R_{ij} = \left(\sum_{f=1}^{F} p_f q_{f_i}\right)^{k_i} \frac{\mathcal{B}\beta}{\sum_{j \in u} x_{ij}} \log(1 + \gamma_{ij}) \qquad (8)$$

In this framework, the ratio of signal to interference-noise can be computed via (11)

$$\gamma_{ij} = \frac{P_i h_{ij}}{\sum_{i' \in \mathcal{B}, i' \neq i} P_i h_{i'j} + \sigma^2} \qquad (11)$$

In this formulation, $h_{ij}$ and $h_{i'j}$ indicate the main channel gain and the interfering channel gain respectively, $B$ denotes the frequency bandwidth. $\sigma^2$ is also a noise figure. We can consider the goal function equivalent to minimization of the applied grid power. Consequently, we have the goal function as the following.

**P1:** $$\max_{q,x,P,\varepsilon,G} \sum_{i \in \mathcal{B}} \sum_{j \in u} x_{ij} \mu_{ij} - \eta \sum_{i \in \mathcal{B}} G_i$$

*s. t.* $C1: \sum_{i \in \mathcal{B}} x_{ij} \gamma_{ij} \geq \gamma_{min}, \forall j \in u,$

$C2: \sum_{i \in \mathcal{B}} x_{jm} = 1, \forall j \in u,$

$C3: P_i < G_i + \beta \sum_{i' \in \mathcal{B}, i' \neq i} \varepsilon_{i'i} - \sum_{i' \in \mathcal{B}, i' \neq i} \varepsilon_{ii'} + E_i, \forall i \in \mathcal{B},$

$C4: \sum_{f=1}^{F} q f_i \leq L_i, \forall i \in \mathcal{B}, f \in \mathcal{F},$

$C5: 0 \leq q_{f_i} \leq 1, \forall f \in \mathcal{F}, \forall i \in \mathcal{B},$

$C6: x_{ij} \in \{0,1\}, \forall i, \forall j \in u,$

$C7: G_i \geq 0, \varepsilon_{ii'} \geq 0, \forall i \in \mathcal{B},$

$C8: 0 \leq P_i \leq P_{max}^i, \forall i \in \mathcal{B},$

In which, $\mathbf{q}=[q_{f_i}]$, $\mathbf{X}=[x_{ij}]$, $\mathbf{P}=[P_i]$, $\boldsymbol{\varepsilon}=[\boldsymbol{\varepsilon}_{il'}]$, $\mathbf{G}=[G_i]$, $\gamma_{min}$ illustrates the $SINR_{min}$ to guarantee the reliability of the connection between nodes and the base station. Also, $\eta$ represents a weighting factor for evaluation of the power efficiency index. The multi hop strategy of backhauling in the presented mobile sensor network and the configuration of the network has been exhibited in Figure 4.

Figure 4 shows a simple network structure to illustrate the operation process of the proposed connectivity approach.

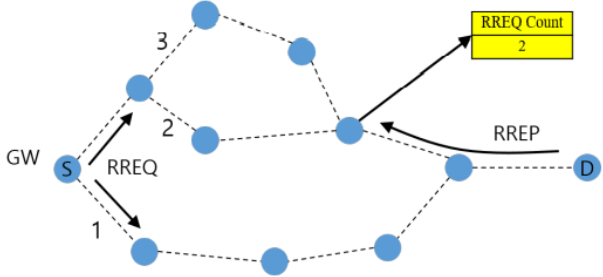

Fig. 4. Operation process of the proposed connectivity approach

In order to improve the reliability of the proposed model, we applied a path repair process with the branch node-based routing algorithm which is shown in Figure 8.

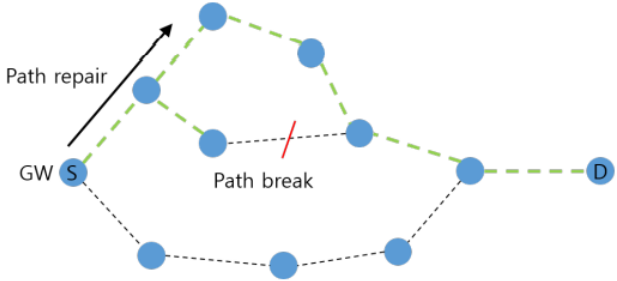

Fig. 5. Reliable distributed transmission model.

In this paper, we present the cost of the $L$ link, which has been optimized for multi-hop paths that minimize the average power consumption of the mobile sensor network. Our distributed cross layer routing protocol uses a connection cost that can be added to the final goal function. Using simulations, the paper shows that the cost of our link is such that it maintains the anonymity of the base station while reducing the communication energy consumption.

The problem of association and power allocation in this approach may be modelled as problem P2 which itself can be considered as the optimal solution for the primary problem P1. Taking $k_i = \sum_{j \in \mathcal{U}} x_{ij}$, into account, this problem is expressed as the following.

$$
\begin{aligned}
\text{P2:} \quad \max_{q,x,P,\varepsilon,G} & \sum_{i \in \mathcal{B}} \sum_{j \in \mathcal{U}} x_{ij} log(c_{ij}) \\
& + \sum_{i \in \mathcal{B}} k_i^2 log\left(\sum_{f=1}^{F} p_f q f_i\right) \\
& - \sum_{i \in \mathcal{B}} k_i log(k_i) \\
& - \eta \sum_{i \in \mathcal{B}} G_i
\end{aligned}
\tag{13}
$$

$s.t \quad C1, C2, C3, C4, C5, C6, C7, C8,$

$$
C9: \sum_{j \in \mathcal{U}} x_{ij} = k_i, \forall i,
$$

where, $c_{ij} = Blog(1 + \gamma_{ij})$.

### D. Data Caching-based User Association Algorithm

In this framework, P2 as a NL mixed integer programming problem is not a convex problem and as Lemma 1 indicated, the sub gradient method will be the best approach to solve it. Taking $\{P, \varepsilon, G\}$ into account, the sensor association problem will be mathematically modeled as follows.

$$
\begin{aligned}
\text{P2.1:} \quad \max_{q,x} & \sum_{i \in \mathcal{B}} \sum_{j \in \mathcal{U}} x_{ij} log(c_{ij}) \\
& + \sum_{i \in \mathcal{B}} k_i^2 log\left(\sum_{f=1}^{F} p_f q f_i\right) \\
& - \sum_{i \in \mathcal{B}} k_i log(k_i)
\end{aligned}
\tag{14}
$$

$s.t. \quad C1, C2, C4, C5, C6, C9$

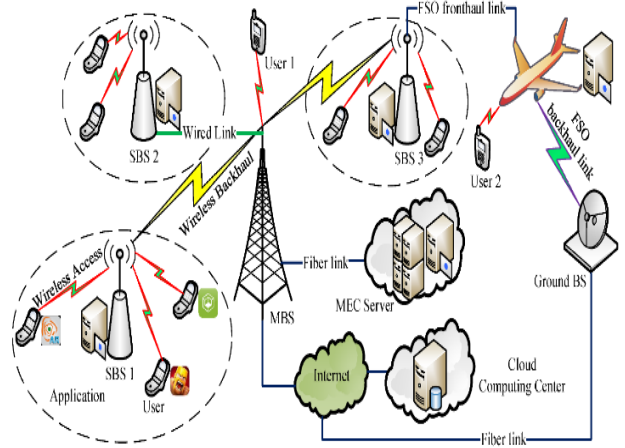

Fig. 4. Beamforming model in the mobile sensor architecture

Lemma 1: considering $p_{(1)} \geq \cdots \geq p_{(f)} \geq \cdots \geq p_{(F)}$ as the probability of demanded payload $(f)$, the optimal solution for P2.1 is achievable as the following.

$$
q_{f_i}^* = \begin{cases} 1, & f_i = (1), \dots, (L_i) \\ 0, & otherwise \end{cases} , \quad \forall i \in \mathcal{B}.
\tag{15}
$$

Proof 1: as mentioned before, P2.1 shows that obtaining the optimal value for $\sum_{f=1}^{F} p_f q_f$ is the target in which, the demanded payload is itself consists of $L_i$ sections $\mathcal{F}_l (l = 1, \ldots, L_i)$, and the probability $\mathcal{F}_l$ is more than $\mathcal{F}_{l+1}$. Therefore, we have:

$$\sum_{(f) \in \mathcal{F}_l} q_{(f)i}^l = 1, \sum_{l=1}^{L_i} q_{(f)i}^l = q_{(f)i} \ and \ \bigcup_{l=L_i} \mathcal{F}_l = \mathcal{F}$$

Also,

$$\sum_{f=1}^{F} p_f q_{fi} = \sum_{l=1}^{L_i} \sum_{(f) \in \mathcal{F}l} p_{(f)} q_{(f)i}^l \le \sum_{l=1}^{L_i} p_{(l)} \left( \sum_{(f) \in \mathcal{F}l} q_{(f)i}^l \right)$$

$$\Rightarrow \sum_{f=1}^{F} p_f q_{f_i} \le \sum_{l=1}^{L_i} p_{(l)},$$

Consequently, according to (15), this theory is confirmed. So, we can conclude that

$$\overline{P}2.1: \quad \begin{aligned} \max_x & \sum_{i \in \mathcal{B}} \sum_{j \in u} x_{ij} log(c_{ij}) + \sum_{i \in \mathcal{B}} k_i^2 log\left( \sum_{f=1}^{L_i} p_{(f)} \right) \\ & - \sum_{i \in \mathcal{B}} k_i log(k_i) \end{aligned} \tag{16}$$

*s. t.* C1, C2, C6, C9.

For simplicity of process to obtain the best solution for $\overline{P}2.1$ as a combination of several sub-problems, we work on its dual problem. Therefore, the target function should be reformulated as follows:

$$\mathcal{L}(x, k, \mu, \nu) = \sum_{i \in \mathcal{B}} \sum_{j \in u} x_{ij} log(c_{ij}) + \sum_{i \in \mathcal{B}} k_i^2 log\left( \sum_{f=1}^{L_i} p_{(f)} \right)$$
$$- \sum_{i \in \mathcal{B}} k_i log(k_i) - \sum_{j \in u} \mu_j \left( \gamma_{min} - \sum_{i \in \mathcal{B}} x_{ij} \gamma_{uj} \right) -$$
$$\sum_{i \in \mathcal{B}} \nu_i \left( \sum_{i \in u} x_{ij} - k_i \right), \quad (17)$$

Where in this formulation, $\nu = [\nu_i], k = [k_i]$ and $\mu = [\mu_j]$. It should be noted that $\nu_i$ and $\mu_j$ represent Lagrangian multipliers. In continue, we can define the problem's dual function $\mathcal{D}(.)$ as the following

$$\mathcal{D}(\mu, \nu) = \begin{cases} \max_{x,k} \mathcal{L}(x, k, \mu, \nu) \\ s.t. \ C2, C6. \end{cases} \tag{18}$$

Subsequently, the dual problem of $\overline{P}2.1$ (16) will be formulated as

$$\min_{\mu \ge 0, \nu \ge 0} \mathcal{D}(\mu, \nu). \tag{19}$$

$\mu_j$ and $\nu_i$ are coefficients of the dual problem and solution of the goal function can be obtained as the following steps

$$x_{ij}^* = \begin{cases} 1, & if \quad i = i^* \\ 0, & otherwise \end{cases}, \tag{20}$$

In (20), $i^* = \arg max_i \left( \log(c_{ij}) + \mu_j \gamma_{ij} - \nu_i \right)$.

Considering $k_i$, the second derivation of the goal function results in

$$\frac{\partial^2 \mathcal{L}}{\partial k_i^2} = 2log\left( \sum_{f=1}^{L_i} p_{(f)} \right) - \frac{1}{k_i}. \tag{21}$$

$$k_i^* = - \frac{W \left( -2log\left( \sum_{f=1}^{L_i} p_{(f)} \right) e^{\nu_i - 1} \right)}{2log \left( \sum_{f=1}^{L_i} p_{(f)} \right)}, \tag{22}$$

As it is obvious, $\sum_{f=1}^{L_i} p_{(f)} \le 1$, so, $\frac{\partial^2 \mathcal{L}}{\partial k_i^2}$ cannot be a positive amount. With setting $\frac{\partial^2 \mathcal{L}}{\partial k_i^2}$ equal to zero, $k_i^*$ is achieved as the optimum degree of $k_i$.

In (22), $W(z)$ shows the Lambert-W factor as a response for $z = we^w$. According to (20), the optimal solution $(\mu^*, \nu^*)$ cannot be achieved by differentializing of $\mathcal{D}(\mu, \nu)$. Therefore, applying the iterative gradient approach will be useful.

$$\mu_j(t+1) = \left[ \mu_j(t) - \delta(t) \left( \sum_{i \in \mathcal{B}} x_{ij}(t) \gamma_{ij} - \gamma_{min} \right) \right]^+, \tag{23}$$

$$\nu_i(t+1) = \left[ \nu_i(t) - \delta(t) \left( k_i(t) - \sum_{j \in u} x_{ij}(t) \right) \right]^+, \tag{24}$$

In this formulation, $x_{ij}(t)$ and $k_i(t)$ can be renewed in an iteration via (20) and (22). The step size was shown by $\delta(t)$ and we have $[a]^+ = \max\{a, 0\}$, $t$ also indicates the iterations quantity.

## 5- Numerical Results

In this section, we present the simulation results that demonstrate the effectiveness of the proposed Multi hop Cooperative Beamforming Mobile Sensor Network (MCB-MSN) approach. For simplicity, we assume that the harvested energy by base station during each time interval is constant. Following the former schemes, we modeled the energy harvest at each base station as the stationary stochastic process. In addition, we assume that the popularity of the content follows the introduced distribution model of [36] and that the contents of the $F$ library have been sorted by popularity. Thus, the probability of the $f^{th}$ demand for popular content is calculated based on [37], where $\alpha$ indicates the skewness of popularity. We compare the performance of the association design and our proposed power optimization based on the signal strength received from the Reference Signal Received Power (RSRP). In the simulation, the sensors randomly move in the macrocellular geographical area and the main simulation parameters have been shown in Table 1.

We studied the power optimization in beamforming-based multi-layer heterogeneous networks. An association algorithm and cooperative power control were proposed to find the optimal data speed in addition to decreasing the network total power utilization. These schemes consider the system security, data speed and energy consumption to be relatively important. Also, in this paper, the effect of the number of sensors and the size of the cache was also investigated.

### A- Simulation Environment

We evaluate the effectiveness of our approach using proprietary computer simulation with the Monte Carlo method. We analyze the results of implementing the proposed approach in MATLAB 2019 and CVX tool of Python programming language. The experiment environment was considered as a multi-layer heterogeneous system with a number of small cells within a micro layer.

The confidence interval of the results is 90%. The $S_U$ sensor nodes are evenly distributed on the grid at $880 \times 880 \ m^2$. The base station is fixed without any mobility. But activated cells are able to submit packets to the base station in cell 35 via multihop paths. This procedure divides the target network into a grid of 36 cells measuring $167 \times 167 m^2$. We considered the sensitivity of each network node to receive a signal equal to -100 dBm, which is the common value in mobile sensor networks and the signal-to-noise (SNR) required a function which are able to estimate the base CSI and $P_T^{S_i, D_i}$ required to reach the next hop destination by observing the signal to noise (SNR). The maximum transmission power of each node is limited to 30 dBm. Distributed beamforming is used for each hop and occurs each time when $|L_A| > 1$.

We measure the performance of the protocol considering the context of base station anonymity (i.e., reducing the belief that the cell contains base station $B(u = 35)$). To assess MCB-MSN from the energy efficiency point of view, we compared it with three other schemes of mobile sensor networks: Fixed Power Allocation (FPA), Random Power Allocation (RPA) and Cooperative NOMA Simultaneous Wireless Information and Power Transfer (CN-SWIPT) [38].

### B- Numerical Results

Figure 5 compares the average throughput of the proposed approach, MCB-MSN and CN-SWIPT scheme with equal maximum transmission power. Based on this figure, it is obvious that the average sum data rate increases with increasing the signal to noise ratio. It can also be seen that the MCB-MSN algorithm performs much better than other algorithms in terms of higher data rate. Because the MCB-MSN algorithm has the required flexibility to dedicate resources to the network entities.

This trend decreases slightly with an increase in N, because the algorithm reduces the throughput available to each of the nodes. In contrast, the demand for the throughput of each node is the same in all random power allocation, equal power allocation, CN-SWIPT algorithms, because they all provide the minimum throughput for each N. The throughput decreases exponentially with increasing number of N. Because with increasing N, the demand for data rate decreases. Because, the same MTP is shared equally between the nodes.

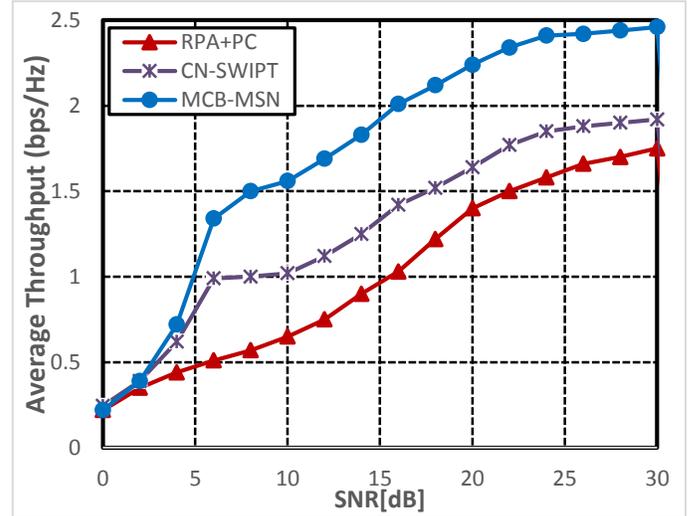

Fig. 5. Average throughput *vs.* signal to noise ratio

Based on the achieved results in Figure 6, the average sum-rate increases almost linearly with increasing N in the MCB-MSN algorithm. While all three other two algorithms, CN-SWIPT and RPA/PC show slight improvement

**Table 1.** Main implementation factors

| Parameter | Value |
|---|---|
| Configuration of the Network | Mobile Network, X-sectored BSs |
| sensor distribution model | uniform (U) and hotspot (Hs), |
| transmit backoff | 1.5 dB |
| Base Station MTP | 43 dBm |
| Codec strategy | Adaptive multi-rate |
| **Rx loss** & **Tx loss** | 3 dB |
| Propagation model | Okumura-Hata |
| Fairness Index | Security/ Throughput |
| Upper bound of iteration | 2000 |
| **$L_{margin}$** | 5 dBm |

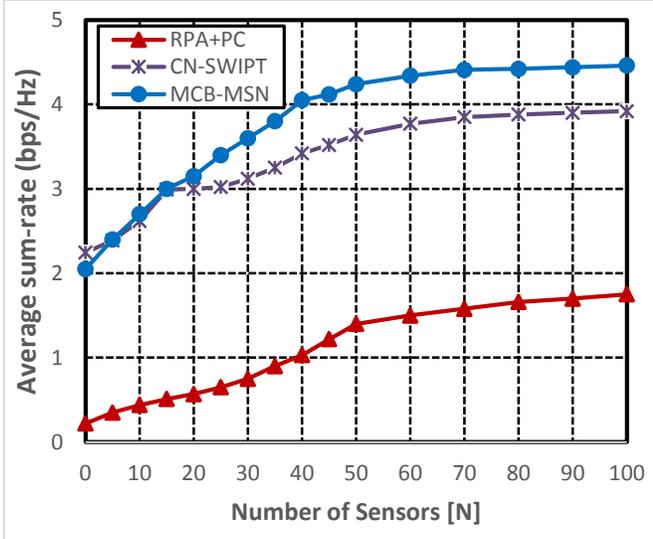

Fig. 6. Average sum rate *vs.* number of sensors

Figure 7 shows the capacity of backhaul links and their average traffic (link usage in percentage) for random power allocation, MSB-MSN and CN-SWIPT. Based on this figure, it can be seen that the MCB-MSN algorithm has the best performance in terms of load balancing and link usage and capacity. So, it has the highest possible efficiency in using backhaul links. Also, the high capacity of backhaul links reduces the potential for the backhaul link to be trapped in the bottleneck while sending the traffic flow to the central network.

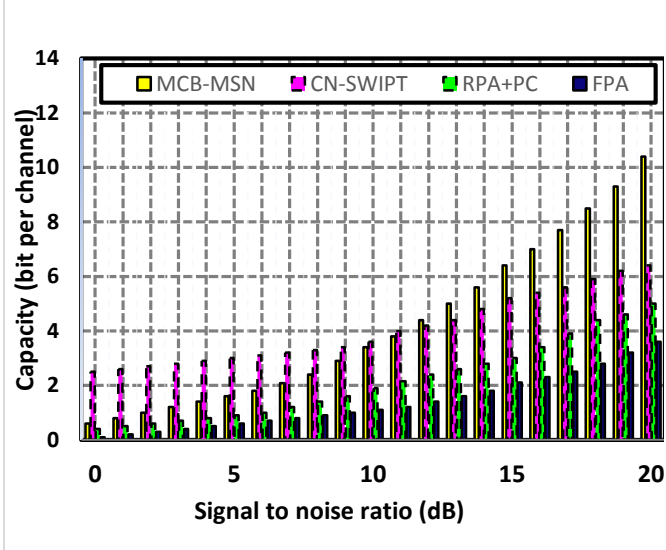

Fig. 7. Backhaul links capacity *vs.* signal to noise ratio

Figure 8 shows the performance of MCB-MSN and CN-SWIPT algorithms according to different numbers of nodes. As can be seen from the graph, energy efficiency is obtained when broader constraints on the total number of nodes are considered. Because, the feasible range of the problem increases and the algorithm has more freedom to maximize the throughput and minimize the energy consumption.

But as the upper bound of demand decreases, the feasible range becomes narrower and energy efficiency decreases. Further lowering the upper bound to $ymin = ymax = Cue$ (where $cue$ is the user equipment demand that must be met) results in identical efficiency of both the MCB-MSN and CN-SWIPT algorithms. Therefore, the MCB-MSN algorithm, which uses dynamic power optimization, performs better than the CN-SWIPT, which has strict constraints procedure.

Figure 9 shows that the MCB-MSN algorithm uses power sources better than other algorithms. Increasing the MTP to saturation increases the energy efficiency. After reaching saturation, increasing MTP does not affect the energy efficiency. Increasing MTP in Equivalent Power Allocation (EPA), Random Allocation (RA) and CN-SWIPT algorithms does not improve energy efficiency. In this case, the performance of these algorithms is slightly worse. In the form of 1000 independent simulations, we investigated how many times each algorithm successfully calculates the solution. Our criterion is actually the possible values used to summarize the result of each simulation result in CVX. CVX as a linear programming method is a powerful optimizer for solving iterative problems like the introduced main problem. Such convex-based tools can also be applied to analyze for rapid prototyping of models and algorithms incorporating convex optimization.

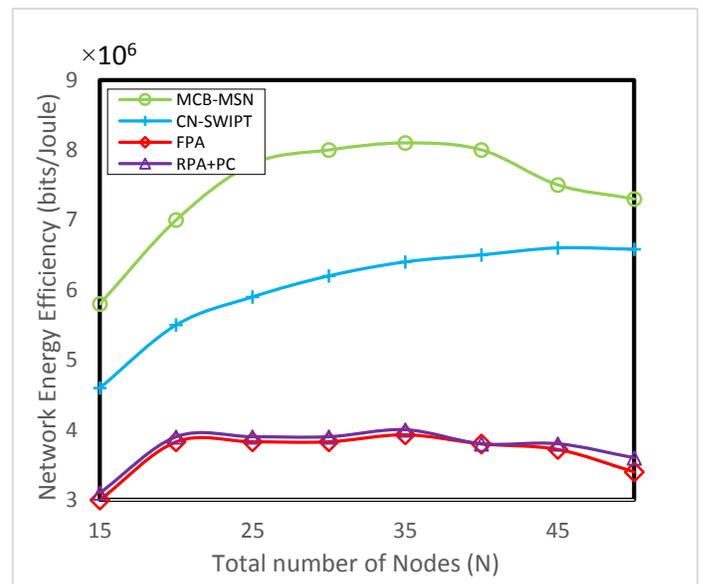

Fig. 8. Network energy efficiency *vs.* total number of nodes

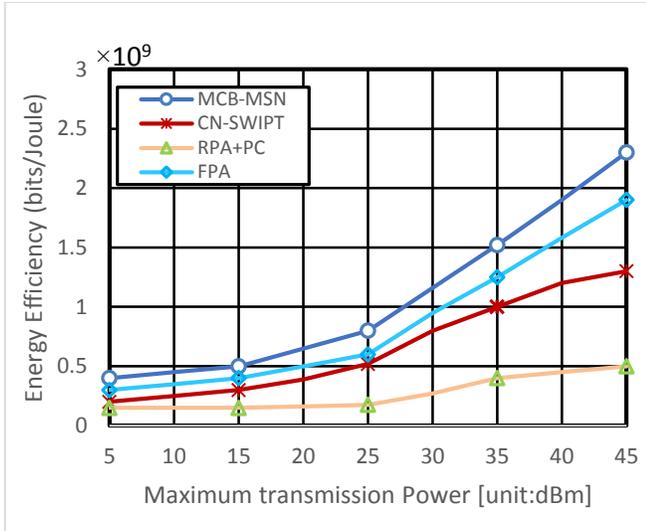

Fig. 9. Network energy efficiency *vs.* maximum transmission power

## 6. Conclusion

This paper presents a novel approach to Multi hop Cooperative Beamforming Mobile Sensor Network (MCB-MSN), which not only increases the anonymity of the base station but also maximizes the network energy efficiency in the distributed beamforming by choosing the routes with higher relay densities. In this paper, when the $\mathcal{L}_i$ link cost of the MCB-MSN algorithm is used, the mobile sensor network maintains its level of anonymity significantly more than in a state where anonymity enhancement techniques are not used. In future studies, more MCB-MSN energy consumption should be evaluated in mobile sensor networks considering non-ideal cooperative beamforming conditions so that information needs to be transmitted frequently. In future, we plan to explore the potential of multi-agent smart queuing in various HetNet scenarios such as privacy-aware recommendation and store cell recommendation. We will also plan to examine how to exploit multi-modal data in the mobile sensor networks to further improve the proposed model.